\documentstyle[12pt,epsf,amsfonts]{article}
\renewcommand{\thefootnote}{\fnsymbol{footnote}}
\begin{document}
\begin{center}
{\large \bf SUPERSYMMETRIC CONSTRUCTION OF\\ EXACTLY SOLVABLE 
POTENTIALS\\[1mm] AND NON-LINEAR ALGEBRAS}%
\footnote{Talk presented by Georg Junker at the  
VIII International Conference on\newline 
\mbox{}$\quad\;\,$ ``Symmetry Methods in Physics'', 
Dubna, Russia, 28 July -- 2 August, 1997.}
\setcounter{footnote}{0}
\\[5mm]
\renewcommand{\thefootnote}{\arabic{footnote})}
{\bf Georg Junker\footnote{\protect Institut f\"ur Theoretische 
Physik, Universit\"at Erlangen-N\"urnberg, Staudtstr.\ 7,\newline
\mbox{}$\quad\;\,$ D-91058 Erlangen, Germany.}
\quad and \quad
Pinaki Roy}\footnote{Physics and Applied Mathematics Unit,
Indian Statistical Institute, Calcutta 700035, India.}
\end{center}
\begin{abstract}
Using algebraic tools of supersymmetric quantum mechanics we 
construct classes of conditionally exactly solvable potentials being 
the supersymmetric partners of the linear or radial harmonic oscillator.
With the help of the raising and lowering operators of these harmonic 
oscillators and the SUSY operators we construct ladder operators for these 
new conditionally solvable systems. 
It is found that these ladder operators together with the 
Hamilton operator form a non-linear algebra which is of quadratic and cubic 
type for the SUSY partners of the linear and radial harmonic oscillator, 
respectively.
\end{abstract}
\mbox{}

\noindent{\bf 1\quad Introduction, summary and outlook}\\[2mm]
During the last decade supersymmetric (SUSY) quantum mechanics has become an 
important tool in various branches of theoretical physics. In particular, in
quantum mechanical problems SUSY has been found to be a very useful algebraic
tool [1]. For example, the class of exactly solvable quantum systems has been
enlarged by such methods [2]. Quite recently these methods have even been 
extended to the construction of conditionally exactly solvable problems [3], 
where, in addition, it has been shown that these systems have a non-linear
algebraic structure.

It is the aim of this paper to generalize the approach given in [3] to a much 
wider class of conditionally exactly solvable systems being the SUSY partners
of the linear or radial harmonic oscillator. In doing so we will first review
the basic tools of SUSY quantum mechanics [1] which we are going to use. In 
Section 3 we will present in some detail the general construction principle 
previously suggested by us [3]. 
In Section 4 we present the results for the linear 
harmonic oscillator. Section 5 and 6 contain our results on the radial harmonic
oscillator with unbroken and broken SUSY, respectively. 

Besides the construction of conditionally exactly solvable problems we also
analyse their algebraic structure, which turns out to be uniquely characterized
be their SUSY partner. That is, for the SUSY partners of the linear oscillator 
we obtain a quadratic algebra and for the radial oscillator (unbroken as well
as broken SUSY) we find a cubic algebra.

As becomes clear from our general method in Section 3, the present approach 
can also be applied to other shape-invariant SUSY systems such as the radial 
hydrogen atom, Morse or P\"oschl-Teller oscillator. Another application is in
the construction of exactly solvable drift potentials associated with the
Fokker-Planck equation. In fact, for the harmonic oscillator case this has,
in essence, already been done by Hongler and Zheng [4].
\\[2mm]

\noindent{\bf 2 \quad Supersymmetric quantum mechanics}\\[2mm]
Witten's model of supersymmetric quantum mechanics consists of a pair of
standard Schr\"odinger Hamiltonians
\begin{equation}
H_\pm=-\frac{1}{2}\,\frac{d^2}{dx^2}+V_\pm(x)
\end{equation}
acting on the Hilbert space ${\cal H}$ of square integrable functions on the 
configuration space $M$, which we will assume to be the real line in the case
of the linear harmonic oscillator, ${\cal H}=L^2({\mathbb R})$, 
or the positive half line in the case of the radial harmonic oscillator,
${\cal H}=\{\psi\in L^2({\mathbb R}^+)| \psi(0)=0\}$.
The so-called SUSY partner potentials 
\begin{equation}
V_\pm(x)= \frac{1}{2}\Bigl(W^2(x)\pm W'(x)\Bigr)
\end{equation}
are given by the SUSY potential $W:M\to{\mathbb R}$ and its derivative 
$W'=dW/dx$. In terms of the SUSY operators
\begin{equation}
A=\frac{1}{\sqrt{2}}\left(\frac{d}{dx}+W(x)\right)\;,\quad
A^\dagger=\frac{1}{\sqrt{2}}\left(-\frac{d}{dx}+W(x)\right)
\label{A}
\end{equation}
the SUSY partner Hamiltonians read $H_+=AA^\dagger\geq 0$ and 
$H_-=A^\dagger A\geq 0$. 

With the help of the operators (\ref{A})
it is easy to show that $H_+$ and $H_-$ are essentially isospectral. 
To be more explicit, let us
denote the eigenfunctions and eigenvalues of $H_\pm$ by $\psi^\pm_n$ and 
$E^\pm_n$, respectively. That is,
\begin{equation}
H_\pm\psi^\pm_n(x)=E^\pm_n\psi^\pm_n(x)\;,\quad n=0,1,2,\ldots\;.
\end{equation}
In the case of unbroken SUSY (here we will use the convention [1] that the 
zero-energy eigenstate of the SUSY system belongs to $H_-$) 
we have for the ground state of $H_-$ the relations
\begin{equation}
E_0^-=0\;,\quad \psi_0^-(x)=C\exp\left\{-\int dx\, W(x)\right\}\in {\cal H}
\label{SUSYGS}
\end{equation}
with $C$ being a proper normalization constant. The remaining spectrum of $H_-$
coincides with the complete spectrum of $H_+$ and the corresponding 
eigenfunctions are related by SUSY transformations:
\begin{equation}
\begin{array}{ll}
E_{n+1}^-=E_n^+>0\;, \quad
 &\psi_{n+1}^-(x)=(E_n^+)^{-1/2}A^\dagger\psi_n^+(x)\;,\\
 &\psi_{n}^+(x)=(E_{n+1}^-)^{-1/2}A\psi_{n+1}^-(x)\;.
\label{SUSY}
\end{array}
\end{equation}

In the case of broken SUSY $H_+$ and $H_-$ are strictly isospectral and the 
eigenfunctions are also related by SUSY transformations:
\begin{equation}
\begin{array}{ll}
E_{n}^-=E_n^+>0\;,\quad 
 &\psi_{n}^-(x)=(E_n^+)^{-1/2}A^\dagger\psi_n^+(x)\;,\\
 &\psi_{n}^+(x)=(E_{n}^-)^{-1/2}A\psi_{n}^-(x)\;.
\label{BSUSY}
\end{array}
\end{equation}
Though above relations (\ref{SUSY}) and (\ref{BSUSY}) are also valid in the 
cases of continuous spectra, we 
consider in this paper only systems having a purely discrete spectrum.

With the help of the relations (\ref{SUSYGS}) and (\ref{SUSY}) or 
(\ref{BSUSY}) it is obvious that
knowing the spectral properties of, say, $H_+$ one immediately obtains the 
complete spectral properties of the SUSY partner Hamiltonian $H_-$. It is this 
fact which is our basis for the construction of (conditionally) exactly 
solvable potentials, by which we mean that the eigenvalues and eigenfunctions
of the corresponding Schr\"odinger Hamilton\-ian can be given in an explicit
closed form (under certain conditions obeyed by the potential parameters). 
Furthermore, the SUSY operators (\ref{A}) also allow to 
construct from known ladder operators of $H_+$ the corresponding ladder
operators for $H_-$ which turn out to closed a non-linear algebra.\\[2mm]

\noindent{\bf 3 \quad Construction of exactly solvable potentials}\\[2mm]
In this section we present our basic idea for the construction of 
(conditionally) exactly solvable potentials. As already anticipated in the 
last section, the basic idea is to choose the SUSY potential $W$ such that
the partner potential $V_+$ becomes one of the well-known exactly solvable 
ones, that is, the eigenvalue problem for the corresponding 
Hamiltonian $H_+$ is exactly solvable. In this way we can eventually find
(through a proper choice of $W$) new 
partner potentials which are also exactly solvable. That is, the spectral 
properties for $H_-$ are obtainable via the SUSY transformations 
(\ref{SUSY}) or (\ref{BSUSY}).

In order to find an appropriate class of SUSY potentials we make the ansatz [3]
\begin{equation}
W(x)=\Phi(x)+f(x)
\end{equation}
where $\Phi$ is a so-called shape-invariant SUSY potential [1], that is, for 
$f\equiv 0$ the corresponding partner potentials $V_\pm$ belong to a known 
class of exactly solvable ones. For a non-vanishing $f$ we have
\begin{equation}
\textstyle
V_+(x)=\frac{1}{2}\Bigl[\Phi^2(x)+\Phi'(x)+f^2(x)+2\Phi(x)f(x)+f'(x)\Bigr]\;.
\end{equation}
If we now choose $f$ such that it obeys the following generalized Riccati 
equation
\begin{equation}
f^2(x)+2\Phi(x)f(x)+f'(x)=2(\varepsilon-1)\;,
\label{DGLf}
\end{equation}
at least under certain conditions on the parameters contained in $\Phi$ 
and certain
values of $\varepsilon\in{\mathbb R}$, than the two partner potentials read
\begin{eqnarray}
V_+(x)&=&\textstyle\frac{1}{2}\Phi^2(x)+\frac{1}{2}\Phi'(x)+\varepsilon-1\;,
\label{V+}\\
V_-(x)&=&\textstyle\frac{1}{2}\Phi^2(x)-\frac{1}{2}\Phi'(x)-f'(x)+
\varepsilon-1\;.
\end{eqnarray}
Clearly, the potential $V_+$ is by construction shape-invariant and therefore
exactly solvable. Via the SUSY transformation we can now also solve the 
eigenvalue problem for $H_-$ associated with the above $V_-$ which, due to
our assumption that the potential parameters had to take certain values, is
sometimes called a conditionally exactly solvable potential [5]. 
A first and obvious
condition on the parameter $\varepsilon$ is that it has to be large enough in
order to give rise to a strictly positive Hamiltonian $H_+>0$. If this would 
not be the case, than the SUSY transformations would lead to ``wavefunctions''
which are not in the Hilbert space ${\cal H}$. This, for example, may happen if
the solution $f$ of (\ref{DGLf}) contains a singularity in the 
configuration space $M$. Note that the SUSY operators (\ref{A}) are given by
\begin{equation}
A=\frac{1}{\sqrt{2}}\left(\frac{d}{dx}+\Phi(x)+f(x)\right)\;,\quad
A^\dagger=\frac{1}{\sqrt{2}}\left(-\frac{d}{dx}+\Phi(x)+f(x)\right)
\label{A'}
\end{equation}
and should leave the Hilbert space invariant, $A:{\cal H}\to{\cal H}$,
$A^\dagger:{\cal H}\to{\cal H}$.

In order to search for such regular solutions of (\ref{DGLf}) we linearize it
by setting $f(x)=u'(x)/u(x)$, which leads to an ordinary, homogeneous and
linear second-order differential equation:
\begin{equation}
u''(x)+2\,\Phi(x)\,u'(x)+2(1-\varepsilon)u(x)=0\;.
\label{DGLu}
\end{equation}
In terms of $u$ the conditionally exactly solvable potential than reads
\begin{equation}
V_-(x)=\frac{1}{2}\,\Phi^2(x)-\frac{1}{2}\,\Phi'(x)+\frac{u'(x)}{u(x)}\left(
2\,\Phi(x)+\frac{u'(x)}{u(x)}\right) -\varepsilon +1\;.
\label{V-}
\end{equation}
The regularity condition on $f$ now amounts to obtain the most general 
solutions of (\ref{DGLu}) which is (without loss of generality) strictly 
positive on $M$. This is equivalent
to require that $V_-$ does not have any additional singularities besides
that of $V_+$. The latter may only exist at $x=0$ for the case 
$M={\mathbb R}^+$.

In the following we will consider three examples corresponding to the linear
harmonic oscillator with unbroken SUSY  and the radial harmonic oscillator
with unbroken as well as broken SUSY. For these systems we also know how to
construct ladder operators which close a linear algebra. With the 
help of the SUSY operators (\ref{A'}) we are then able to obtain ladder 
operators for
the conditionally exactly solvable system $H_-$ which turn out to close a 
non-linear algebra.\\[2mm]

\noindent{\bf 4 \quad The linear harmonic oscillator}\\[2mm]
As a first example we will consider the SUSY potential of the linear
harmonic oscillator on the real line $M={\mathbb R}$:
\begin{equation}
\Phi(x)=x
\label{Phi1}
\end{equation}
It is straightforward to verify that in this case the potential (\ref{V+}) is
indeed that of the linear harmonic oscillator
\begin{equation}
V_+(x)=\textstyle\frac{1}{2}\,x^2+\varepsilon-\frac{1}{2}
\label{V+1}
\end{equation}
whose energy eigenvalues and eigenfunctions are
\begin{equation}
E_n^+=n+\varepsilon\;,\quad
\psi_n^+(x)=\left[\sqrt{\pi}\,2^nn!\right]^{-1/2}H_n(x)\exp\{-x^2/2\}\;,
\end{equation}
where $H_n$ denotes the Hermite polynomial of order $n=0,1,2,\ldots$. As we
require a strictly positive spectrum for $H_+$ we arrive at a first condition 
on the parameter $\varepsilon$ which reads $\varepsilon>0$. 

Let us now turn to the solution of (\ref{DGLu}) with the linear SUSY potential
(\ref{Phi1}). With the substitution $z=-x^2$ this differential equation can
be transformed into that of the confluent hypergeometric function and thus the
most general solution reads ($\alpha,\beta\in{\mathbb R}$ are two additional
system parameters) 
\begin{equation}
\begin{array}{ll}
u(x)&=\alpha\,_1F_1\left(\frac{1-\varepsilon}{2},\frac{1}{2},-x^2\right) + 
\beta\,x\,_1F_1\left(\frac{2-\varepsilon}{2},\frac{3}{2},-x^2\right)\\
&=e^{-x^2}\Bigl[
\alpha\,_1F_1\left(\frac{\varepsilon}{2},\frac{1}{2},x^2\right)+
\beta\,x\,_1F_1\left(\frac{1+\varepsilon}{2},\frac{3}{2},x^2\right)\Bigr]\;.
\end{array}
\label{u1}
\end{equation}
As we are searching for strictly positive solutions the real parameter $\alpha$
must not vanish and thus can be set to unity without loss of generality. In 
addition the real parameter $\beta$ has to obey the inequality 
$|\beta|<2\,\Gamma(\frac{1+\varepsilon}{2})/\Gamma(\frac{\varepsilon}{2})$
which follows from the positivity condition $u>0$ via the asymptotic form
\begin{equation}
u(x)=x^{\varepsilon-1}\left(\frac{\Gamma(1/2)}{\Gamma(\varepsilon/2)}+\beta\,
\frac{\Gamma(3/2)}{\Gamma(\frac{1+\varepsilon}{2})}\right)
\Bigl[1+O(1/x)\Bigr]\;.
\label{uasym1}
\end{equation}
Note that for $\beta=0$ the positivity requirement on $u$ leads to 
$\varepsilon>0$, a condition already obtained above from the positivity 
of $H_+$.
Under these conditions the potential (\ref{V-}) is given by
\begin{equation}
V_-(x)=\frac{1}{2}\,x^2-\varepsilon+\frac{1}{2}+\frac{u'(x)}{u(x)}\left(
2\,x+\frac{u'(x)}{u(x)}\right)
\label{V-1}
\end{equation}
which is now a conditionally exactly solvable potential. A plot of this
potential for $0<\varepsilon \leq 3$ and $\beta=0$ is given in Figure 1.
For small $\varepsilon$ and $\beta=0$ the potential $V_-$ exhibits two deep 
and one shallow minimum which is located at the origin. In fact, the parameter
$\varepsilon$ is the tunneling splitting due to the tunnel effect between the
two deep minima.
For large values of $\varepsilon$ the shallow minimum at the center $x=0$
becomes deeper and the other two minima, which are 
symmetrically located about the origin, disappear. 
For non-vanishing $\beta$ the basic structure of $V_-$ is the same but now 
it is no longer symmetric about $x=0$.
\begin{figure}[t]
\vspace{90mm}
\caption{The family (\ref{V-1}) of SUSY 
partner potentials corresponding to the
linear harmonic oscillator potential (\ref{V+1}). Here we have shown only
the symmetric case $\beta=0$.}
\end{figure}

The groundstate energy eigenvalue and eigenfunction of $H_-$ for
the above potential (\ref{V-1}) are given by
\begin{equation}
E_0^-=0\;,\quad\psi_0^-(x)=\frac{C}{u(x)}\exp\{-x^2/2\}\;.
\label{LHOE0}
\end{equation}
Note that because of (\ref{uasym1}) the above groundstate wavefunction is
square integrable and therefore SUSY is unbroken. The remaining spectral
properties of $H_-$ follow from those of $H_+$ via the SUSY transformation
(\ref{SUSY}):
\begin{equation}
E_{n+1}^-=n+\varepsilon \;,\quad
\psi_{n+1}^-(x)=
\frac{\exp\{-x^2/2\}}{\left[\sqrt{\pi}\,2^{n+1}n!(n+\varepsilon)\right]^{1/2}}
\left(H_{n+1}(x)+H_n(x)\frac{u'(x)}{u(x)}\right).
\label{LHOEn}
\end{equation}
Let us also remark that for $\beta=0$ and an odd integer $\varepsilon=2N+1>0$ 
the solution
(\ref{u1}) becomes a polynomial in $x^2$ of degree $N$ with no real zeros,
that is, $u(x)=(1+g_1x^2)\cdots(1+g_Nx^2)$ with $g_i>0$. These cases, in 
particular for $N=1$ and $2$, have been discussed in [3]. See, however, also [6] 
for a different approach to such cases and their connection to non-linear
superalgebras.

Let us now turn to the construction of ladder operators for $H_-$. In doing
we first recall the well-known ladder operators for the linear harmonic
oscillator $H_+$
\begin{equation}
a=\frac{1}{\sqrt{2}}\left(\frac{d}{dx}+x\right)\;,\quad
a^\dagger=\frac{1}{\sqrt{2}}\left(-\frac{d}{dx}+x\right)\;,
\label{a}
\end{equation}
which obey the linear algebra
\begin{equation}
[H_+,a]=-a\;,\quad [H_+,a^\dagger]=a^\dagger\;,\quad
[a,a^\dagger]=1
\end{equation}
and act on the eigenstates of $H_+$ as follows
\begin{equation}
a\psi_n^+(x)= \sqrt{n}\,\psi_{n-1}^+(x)\;,\quad
a^\dagger\psi_n^+(x)= \sqrt{n+1}\,\psi_{n+1}^+(x)\;.
\end{equation}
With the help of the SUSY operators (\ref{A'}) we can now construct similar
ladder operators [3] for the SUSY partner $H_-$:
\begin{equation}
B=A^\dagger a A\;,\quad B^\dagger=A^\dagger a^\dagger A\;.
\end{equation}
Obviously, these operators act as lowering and raising operators:
\begin{equation}
B\psi_{n+1}^-(x)=\sqrt{E_{n-1}^+nE_{n+1}^-}\,\psi_{n}^-(x)\;,\quad
B^\dagger \psi_{n+1}^-(x)= \sqrt{E_{n+1}^+(n+1)E_{n+1}^-}\,\psi_{n+2}^-(x)\;.
\end{equation}
However, the ground state remains isolated, that is,
$B\psi_{0}^-(x)=0= B^\dagger \psi_{0}^-(x)$.
With these relation one can easily verify that the ladder operators $B$ and 
$B^\dagger $ close together with the Hamiltonian $H_-$ the non-linear algebra
\begin{equation}
[H_-,B]=-B\;,\quad [H_-,B^\dagger ]=B^\dagger\;,\quad 
[B,B^\dagger]=3H_-^2-(2\varepsilon-1)H_-\;,
\end{equation}
which is of quadratic type.
Due to unbroken SUSY, i.e.\ $H_-\psi_{0}^-(x)=0$, this algebra is defined
on the full Hilbert space ${\cal H}=L^2({\mathbb R})$.\\[2mm]

\noindent{\bf 5 \quad The radial harmonic oscillator with unbroken SUSY}\\[2mm]
As a second example we consider the SUSY potential
\begin{equation}
\Phi(x)=x-\frac{\gamma+1}{x}\;,\quad \gamma\geq 0\;,
\label{Phi2}
\end{equation}
which in turn gives rise to the radial harmonic oscillator potential
\begin{equation}
V_+(x)=\frac{x^2}{2}+\frac{(\gamma+1)(\gamma+2)}{2x^2}+\varepsilon-\gamma-
\frac{3}{2}\;.
\label{V+2}
\end{equation}
The energy eigenvalues and eigenfunction of the corresponding Hamiltonian $H_+$
are
\begin{equation}
E_n^+=2n+1+\varepsilon\;,\quad
\displaystyle \psi_n^+(x)=
\left[\frac{2\,n!}{\Gamma(n+\gamma+5/2)}\right]^{1/2}x^{\gamma+2}
L_n^{\gamma+3/2}(x^2)e^{-x^2/2}\,,
\end{equation}
where $L_n^{\nu}$ denotes a Laguerre polynomial of degree $n$ [7]. 
Positivity of $H_+$ leads to the first restriction, $\varepsilon>-1$. 

Let us now consider the
positive solutions of (\ref{DGLu}), which reads
\begin{equation}
\textstyle
u(x)={}_1F_1(\frac{1-\varepsilon}{2},-\gamma-\frac{1}{2},-x^2)
+\beta \,x^{2\gamma+3}
{}_1F_1(2+\gamma-\frac{\varepsilon}{2},\frac{5}{2}+\gamma,-x^2)\;.
\end{equation}
Here we have already set the parameter $\alpha=1$ without loss of generality.
Positivity of the above solution amounts in requiring the following conditions
on the parameters $\beta$, $\gamma$ and $\varepsilon$:
\begin{equation}
0<\frac{\Gamma(-\gamma-\frac{1}{2})}{\Gamma(\varepsilon/2-\gamma-1)}\;,
\quad
|\beta|<\frac{\Gamma(-\gamma-\frac{1}{2})}{\Gamma(\varepsilon/2-\gamma-1)}
\frac{\Gamma(\frac{1+\varepsilon}{2})}{\Gamma(5/2+\gamma)}\;.
\label{cond2}
\end{equation}
The corresponding partner potential reads
\begin{equation}
V_-(x)=\frac{x^2}{2}+\frac{\gamma(\gamma+1)}{2x^2}-\gamma-\varepsilon
+\frac{1}{2}+\frac{u'(x)}{u(x)}
\left(2x-2\frac{\gamma+1}{x}+\frac{u'(x)}{u(x)}\right)
\label{V-2}
\end{equation}
which, because of the above conditions, is now also a conditionally exactly
solvable potential. 
\begin{figure}[t]
\vspace{90mm}
\caption{A family of SUSY partner potentials 
(\ref{V-2}) corresponding to the
radial harmonic oscillator class (\ref{V+2}). Here we have only shown the 
cases $\beta=0$ and $\gamma=1$. Note that because of condition (\ref{cond2})
the allowed ranges of $\varepsilon$ are $0<\varepsilon<2$ and 
$4<\varepsilon<\infty$. For the forbidden regions the figure clearly shows
singularities in $V_-$.}
\end{figure}
In Figure 2 we have shown this potential for $\beta=0$,
$\gamma=1$ and $-1<\varepsilon\leq 8$. Note that for $\varepsilon\leq 0$ and 
$2\leq\varepsilon\leq 4$ the potential (\ref{V-2}) exhibits singularities
as expected because these values of $\varepsilon$ are not allowed for 
$\gamma=1$. As SUSY remains unbroken for all the allowed values of the 
parameters the groundstate energy of the SUSY partner Hamiltonian $H_-$
vanishes and the corresponding eigenstate is obtained from (\ref{SUSYGS}). The 
remaining spectral properties of $H_-$ are found via the SUSY 
transformations (\ref{SUSY}):
\begin{equation}
\begin{array}{l}
\displaystyle
E_0^-=0\;,\quad \quad 
\psi_0^-(x)=\frac{C}{u(x)}\,x^{\gamma+1}e^{-x^2/2}\;,\\
E_{n+1}^-=2n+1+\varepsilon\;,\\
\displaystyle
\psi_{n+1}^-(x)=\frac{1}{\sqrt{4n+2+2\varepsilon}}
\left(-\frac{d}{dx}+x-\frac{\gamma+1}{x}+\frac{u'(x)}{u(x)}\right)\psi_n^+(x)
\;.
\end{array}
\end{equation}

In order to construct the ladder operators for $H_-$ we recall the 
corresponding operators for the radial harmonic oscillator [7] which in essence
are build up from those given in (\ref{a}):
\begin{equation}
c=a^2-\frac{(\gamma+1)(\gamma+2)}{2x^2}\;,\quad
c^\dagger=(a^\dagger)^2-\frac{(\gamma+1)(\gamma+2)}{2x^2}\;.
\end{equation}
These operators act on the eigenstates of $H_+$ as follows
\begin{equation}
\begin{array}{l}
c\,\psi_n^+(x)= -2\sqrt{n(n+\gamma+3/2)}\,\psi_{n-1}^+(x)\;,\\
c^\dagger\,\psi_n^+(x)=-2 \sqrt{(n+1)(n+\gamma+5/2)}\,\psi_{n+1}^+(x)\;.
\end{array}
\end{equation}
and, as in the previous example, close a linear Lie algebra
\begin{equation}
[H_+,c]=-2c\;,\quad [H_+,c^\dagger]=2c^\dagger\;,\quad
[c,c^\dagger]=4(H_++\gamma-\varepsilon+3/2)\;.
\end{equation}
Furthermore, they also allow to construct ladder operators for the 
quantum system characterized by $H_-$:
\begin{equation}
D=A^\dagger c A\;,\quad D^\dagger=A^\dagger c^\dagger A\;.
\end{equation}
These operators act on eigenstates of $H_-$ in the following way:
\begin{equation}
\begin{array}{l}
D\,\psi_{n+1}^-(x)=-2\sqrt{E_{n-1}^+n(n+\gamma+3/2)E_{n+1}^-}\,
\psi_{n}^-(x)\;,\\
D^\dagger\,\psi_{n+1}^-(x)=-2\sqrt{E_{n+1}^+(n+1)(n+\gamma+5/2)E_{n+1}^-}\,
\psi_{n+2}^-(x)\;,\\
D\,\psi_{0}^-(x)=0=D^\dagger\,\psi_{0}^-(x)\;.
\end{array}
\end{equation}
The last line shows that the ground state is again isolated, a fact
due to unbroken SUSY. From the above relations one verifies that these 
operators together with the Hamiltonian close the non-linear algebra
\begin{equation}
\begin{array}{l}
[H_-,D]=-2D\;,\quad [H_-,D^\dagger ]=2D^\dagger\;, \\[1mm]
[D,D^\dagger ]=8H_-^3+12(\gamma-\varepsilon+3/2)H_-^2
-4(2\varepsilon\gamma-\varepsilon^2+3\varepsilon-1)H_-\;,
\label{D2}
\end{array}
\end{equation}
which is of cubic type.\\[2mm]

\noindent{\bf 6 \quad The radial harmonic oscillator with broken SUSY}\\[2mm]
So far we have considered only examples with unbroken SUSY. However, the
radial harmonic oscillator also allows for a broken SUSY. Here in essence the 
second term in (\ref{Phi2}) is opposite in sign. Hence, we consider the
SUSY potential [1]
\begin{equation}
\Phi(x)=x+\frac{\gamma+1}{x}\;,\quad \gamma\geq 0\;,
\label{Phi3}
\end{equation}
which yields the radial harmonic oscillator potential
\begin{equation}
V_+(x)=\frac{x^2}{2}+\frac{\gamma(\gamma+1)}{2x^2}+\varepsilon+\gamma+
\frac{1}{2}\;
\label{V+3}
\end{equation}
and the following spectral properties of the corresponding Hamiltonian $H_+$
\begin{equation}
E_n^+=2n+2\gamma+2+\varepsilon\;,\quad
\psi_n^+(x)=
\left[\frac{2\,n!}{\Gamma(n+\gamma+3/2)}\right]^{1/2}x^{\gamma+1}
L_n^{\gamma+1/2}(x^2)e^{-x^2/2}\,.
\end{equation}
Clearly, we have the condition $-2-2\gamma<\varepsilon$. This condition is
identical with the one obtained from positivity of the solution of (\ref{DGLu})
\begin{equation}
\textstyle
u(x)={}_1F_1(\frac{1-\varepsilon}{2},\gamma+\frac{3}{2},-x^2)\;.
\end{equation}%
\begin{figure}[t]
\vspace{90mm}
\caption{A family of SUSY partner potentials 
(\ref{V-3}) corresponding to the radial harmonic oscillator class (\ref{V+3})
with broken SUSY. Here we have only shown the case $\gamma=1$. 
Note that the allowed range of $\varepsilon$ is $-4<\varepsilon$.}
\end{figure}%
Note that the second linearly independent solution of (\ref{DGLu}) is not 
allowed ($\beta=0$) in order for SUSY to remain broken. The corresponding 
partner potential reads
\begin{equation}
V_-(x)=\frac{x^2}{2}+\frac{(\gamma+1)(\gamma+2)}{2x^2}+\gamma-\varepsilon
+\frac{3}{2}+\frac{u'(x)}{u(x)}
\left(2x+2\frac{\gamma+1}{x}+\frac{u'(x)}{u(x)}\right)
\label{V-3}
\end{equation}
and the spectral properties of the associated $H_-$ are immediately obtained from
(\ref{BSUSY})
\begin{equation}
\begin{array}{l}
E_{n}^-=2n+2\gamma+2+\varepsilon\;,\\
\displaystyle
\psi_{n}^-(x)=\frac{1}{\sqrt{4n+4\gamma+4+2\varepsilon}}
\left(-\frac{d}{dx}+x+\frac{\gamma+1}{x}+\frac{u'(x)}{u(x)}\right)\psi_n^+(x)
\;.
\end{array}
\end{equation}
In Figure 3 we show the potential (\ref{V-3}) for $\gamma=1$ and
$-5\leq\varepsilon\leq 2$. 

As before, we can introduce ladder operators
$D=A^\dagger c A$ and $D^\dagger=A^\dagger c^\dagger A$ which obey the 
non-linear algebra
\begin{equation}
\begin{array}{l}
[H_-,D]=-2D\;,\quad [H_-,D^\dagger ]=2D^\dagger\;, \\[1mm]
[D,D^\dagger ]=8H_-^3-12(\gamma+\varepsilon+1/2)H_-^2
+4(2\varepsilon\gamma+\varepsilon^2+\varepsilon+1)H_-\;.
\end{array}
\end{equation}
This algebra can also be obtained from the unbroken SUSY case (\ref{D2})
by replacing $\gamma$ by $-\gamma-2$. However, in contrast to the unbroken
case, here the ladder operators act on all eigenstates of $H_-$. In other
words, the ground state is not isolated. In fact we have the relation
\begin{equation}
\textstyle
\psi_{n}^-(x)=\left(-\frac{1}{4}\right)^n\left[
n!(\gamma+\frac{3}{2})_n(\gamma+1+\frac{\varepsilon}{2})_n
(\gamma+2+\frac{\varepsilon}{2})_n\right]^{-1/2}
(D^\dagger)^n\psi_{0}^-(x)
\end{equation}
with groundstate wavefunction
\begin{equation}
\psi_{0}^-(x)=\frac{x^{\gamma+1}\exp\{-x^2/2\}}{\sqrt{(2\gamma+\varepsilon+2)
\Gamma(\gamma+\frac{3}{2})}}
\left(2x-\gamma-1+\frac{\gamma+1}{x}+\frac{u'(x)}{u(x)}\right)\;.
\end{equation}
A discussion for the special case $\varepsilon=3$ and arbitrary $\gamma$
is given in [3].\\[2mm]

\noindent{\bf Acknowledgements}\\[2mm]
One of the authors (GJ) would like to thank the organizers of the VIII 
International Conference on ``Symmetry Methods in Physics'' for their kind
invitation to this meeting and the Deutsche Forschungsgemeinschaft for
travel support. The other author (PR) thanks the Deutsche Forschungsgemeinschaft
and the Indian National Science Academy for support.\\[2mm]

\noindent{\bf References}
\renewcommand{\baselinestretch}{0.85}
\small
\begin{enumerate}
\itemsep0mm \parsep-1mm
\item G.\ Junker, {\it Supersymmetric Methods in Quantum and Statistical 
Physics}, Springer, Berlin, 1996.
\item P.B.\ Abraham and H.E.\ Moses, {\it Phys.\ Rev.\ A}, 1980, vol.\ 22, p.\
1333;
B.\ Mielnik, {\it  J.\ Math.\ Phys.}, 1984,  vol.\ 25, p.\ 3387;
M.M.\ Nieto, {\it Phys.\ Lett.}, 1984,  vol.\ 145B, p.\ 208;
D.L.\ Pursey, {\it Phys.\ Rev.\ D}, 1986,  vol.\ 343, p.\ 1048,
P.\ Roy and R.\ Roychoudhury, {\it Z.\ Phys.\ C}, 1986,  vol.\ 31, p.\ 111;
D.\ Zhu, {\it  J.\ Phys.\ A}, 1987,  vol.\ 20, p.\ 4331;
N.A.\ Alves and E.D.\ Filho, {\it J.\ Phys.\ A}, 1988, vol.\ 21, p.\ 1011.
\item G.\ Junker and P.\ Roy, {\it Phys.\ Lett.\ A} , 1997, vol.\ 232, p.\ 155.
\item M.-O.\ Hongler and W.M.\ Zheng, {\it J.\ Stat.\ Phys.}, 1982, vol.\ 29,
p.\ 317; {\it J.\ Math.\ Phys.}, 1983, vol.\ 24, p.\ 336.
\item A.\ de Souza Dutra, {\it Phys.\ Rev.\ A}, 1993, vol.\ 47, p.\ R2435;
R.\ Dutt, A.\ Khare and Y.P.\ Varshni, {\it J.\ Phys.\ A}, 1995, vol.\ 28,
p.\ L107.
\item V.G.\ Bagrov and B.F.\ Samsonov, {\it Teor.\ Mat.\ Fiz.}, 1995, vol.\ 104,
p.\ 356; {\it J.\ Phys.\ A}, 1996, vol.\ 29, p.\ 1011.
\item  A.\ Perelomov, {\it Generalized Coherent States and Their Applications},
Springer, Berlin, 1986.
\end{enumerate}


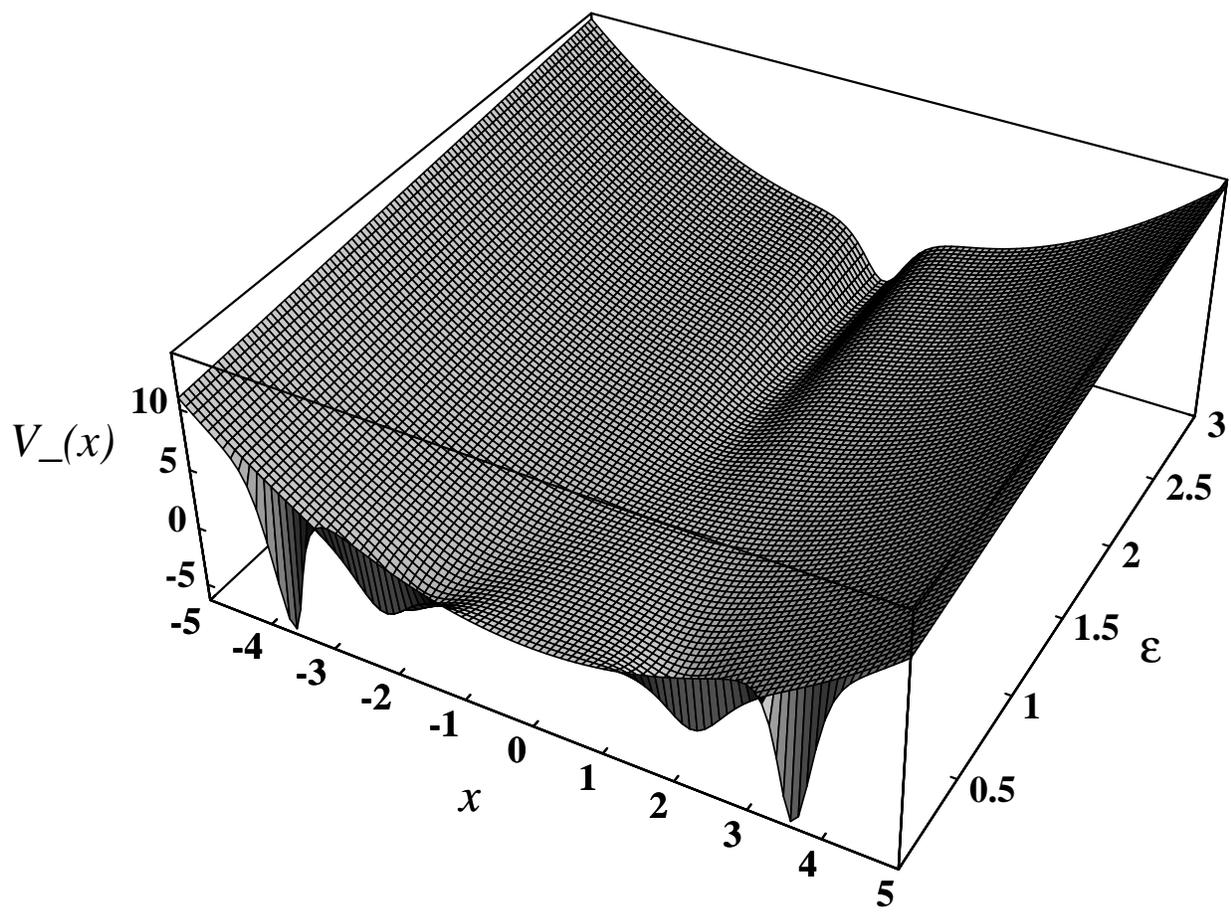
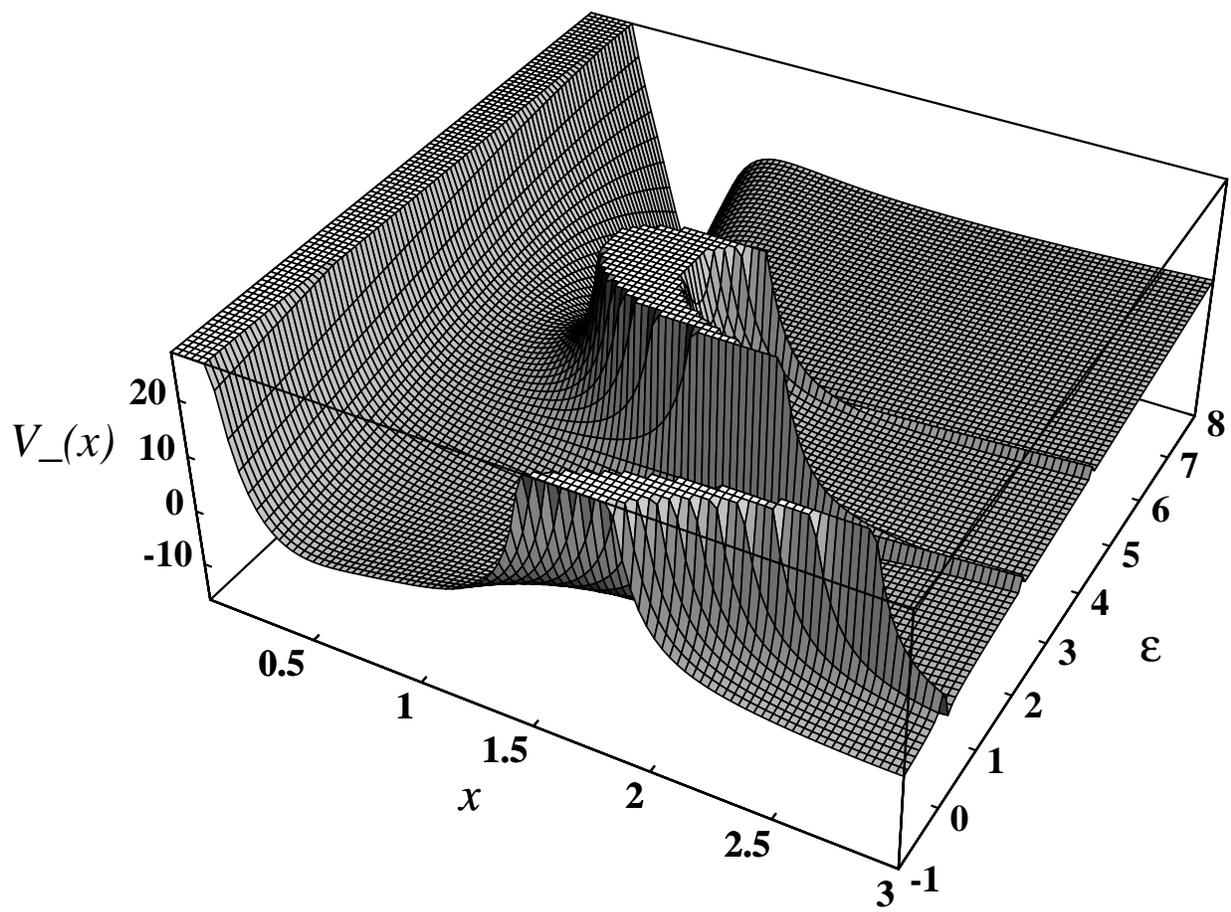
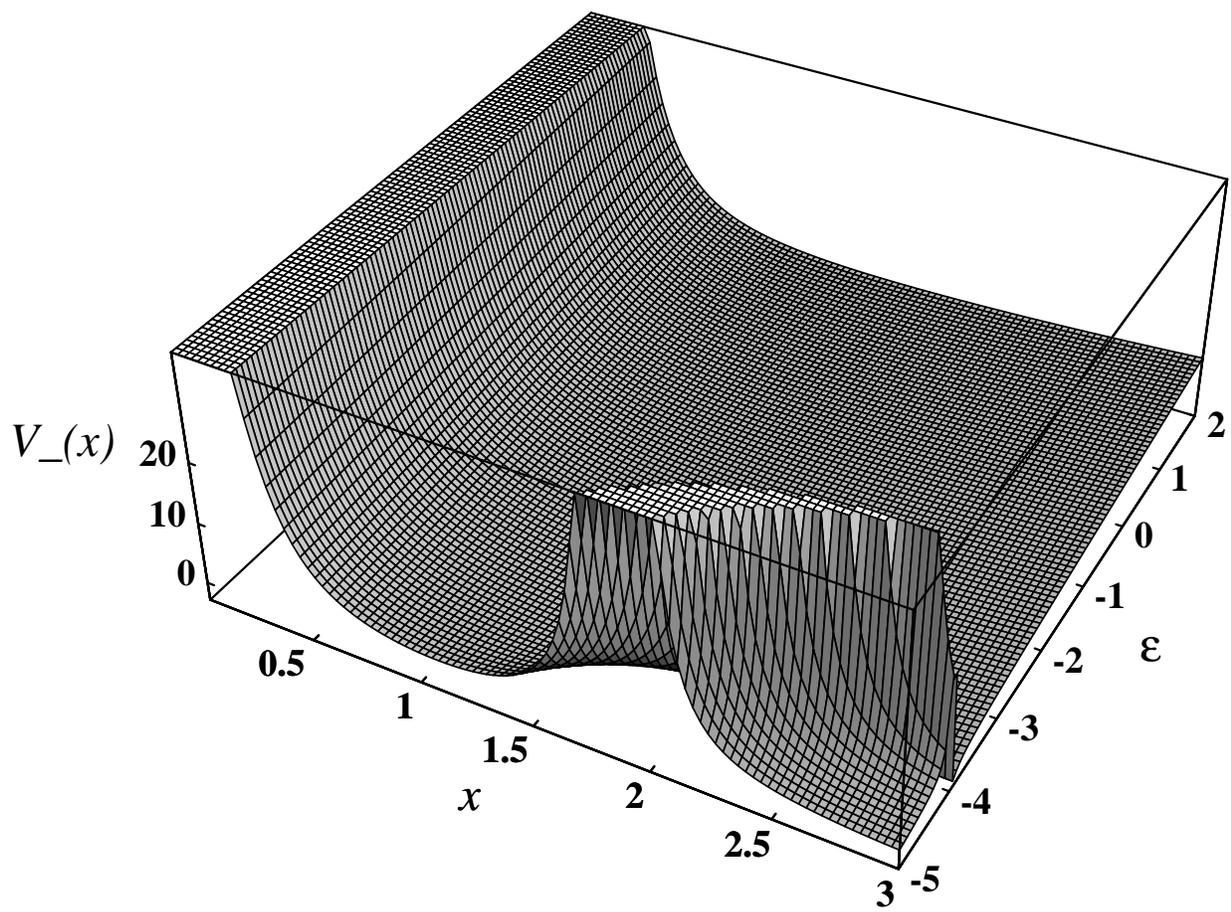
\end{document}